\documentclass{aastex}
\usepackage{apjfonts}
\usepackage{emulateapj5}

\shorttitle{Early Type Galaxies}
\shorttitle{Loeb \& Peebles}

\submitted{Accepted for publication in ApJ}

\newcommand{\beq}{\begin{equation}}
\newcommand{\eeq}{\end{equation}}

\begin{document}

\title{Cosmological Origin of the Stellar Velocity Dispersions\\ 
in Massive Early-Type Galaxies}

\author{Abraham Loeb\altaffilmark{1,3} and P. J. E. 
Peebles\altaffilmark{2}}

\email{loeb@ias.edu; pjep@pupgg.princeton.edu}

\altaffiltext{1}{Institute for Advanced Study, Princeton, NJ 08540}

\altaffiltext{2}{Joseph Henry Laboratories, Princeton University,
Princeton NJ 08544}

\altaffiltext{3}{Guggenheim fellow; on sabbatical leave from the
Astronomy Department, Harvard University, Cambridge MA 02138}

\begin{abstract}
We show that the observed upper bound on the line-of-sight velocity
dispersion of the stars in an early-type galaxy, $\sigma _e\lesssim
400$ km~s$^{-1}$, may have a simple dynamical origin within the
$\Lambda$CDM cosmological model, under two main hypotheses. The first
is that most of the stars now in the luminous parts of a giant
elliptical formed at redshift $z\gtrsim 6$.  Subsequently, the stars
behaved dynamically just as an additional component of the dark
matter. The second hypothesis is that the mass distribution
characteristic of a newly formed dark matter halo forgets such details
of the initial conditions as the stellar ``collisionless matter'' that
was added to the dense parts of earlier generations of halos. We also
assume that the stellar velocity dispersion does not evolve much at
$z\lesssim 6$, because a massive host halo grows mainly by the
addition of material at large radii well away from the stellar core of
the galaxy. These assumptions lead to a predicted number density of
ellipticals as a function of stellar velocity dispersion that is in
promising agreement with the Sloan Digital Sky Survey data.

\end{abstract}


\section{Introduction}
The line-of-sight velocity dispersion of the stars in an elliptical
galaxy with luminosity $L>L_\star$ is typically
$\sigma_e\sim 200$~km~s$^{-1}$, while elliptical or cD galaxies with twice
this velocity dispersion are exceedingly rare. This is equivalent to a
fairly sharp bound on the mass that is gathered within the luminous
parts of the largest galaxies. One might expect this striking effect
has a simple explanation. Our proposal follows a simple path through
the $\Lambda$CDM model for cosmology and structure formation.

We begin in the next section by considering the simple case where
dissipative processes in the baryons are ignored: all matter is
treated as collisionless and initially cold. As discussed in 
\S 2.1, the standard picture for mass clustering in the $\Lambda$CDM
cosmology predicts that in this case the rare extreme mass
concentrations characteristic of the luminous parts of giant
elliptical galaxies have comoving number density as a function of
velocity dispersion that is strikingly similar to what is observed for
these galaxies today.

Reality has to be more complicated than this, because baryons must
dissipatively settle to form stars that make appreciable contributions
to the mass within the effective radii of ellipticals, $R_e$. As we
will discuss, if stellar mass were simply added to the cold dark
matter (CDM) present in these cores the velocity dispersion within the
characteristic effective radius $R_e\sim 10$~kpc would be
unacceptably large. Our proposed remedy invokes two postulates. The
first is that the density profile in a dark matter halo acts as an
attractor or fixed point in the sense of nonlinear dynamics (Syer
\&\ White 1998): the
formation of a new halo tends to erase memory of the conditions
in previous generations of halos, including the distortion caused
by the addition of stellar 
``collisionless matter'' to the central regions. This requires our
second postulate, that the bulk of the stars formed when the mass
concentrations characteristic of the luminous parts of the giant
elliptical galaxies were still being assembled.

We have a measure of when assembly on the scale of the optical
parts of the largest galaxies was close to complete, from the number
density of mass peaks with mass greater than $M_e$ inside a
centered sphere with physical radius $R_e\sim 10$~kpc. At fixed
comoving number density $n(>M_e,t)$, the 
mass $M_e$ increases with increasing time at a redshift $z\sim 10$,
because the dense regions of the halos are still being assembled then, 
while near the present epoch $M_e$ is close to constant, because the
dense central regions of normal galaxies are not much
affected by the ongoing growth of the halo through the addition of
matter at much larger radii. We calculate that the transition is
about at redshift $z_f\simeq 6$. Thus, within this model we must
postulate that the bulk of the stars in a giant elliptical formed
and were assembled into a first approximation to the present-day
galaxy at $z_f\simeq 6$.   

The situation is still more complicated by the evidence that the
density profile within the effective radius of a present-day giant
elliptical differs from the standard estimates of the inner density
profiles of pure cold dark matter halos. This requires yet another
hypothesis, that star formation at low redshift has rearranged the
stellar mass distribution. As we discuss in \S 3 there is 
evidence for modest recent star formation in the central parts of
giant ellipticals, perhaps in part due to the recycling of mass
shed by evolving stars. 

Three points may be of particular interest. First, stars
that form at high redshift behave thereafter dynamically as dark
matter particles. The numerical experiments
reviewed in \S 2.3 suggest the density profiles in the subsequent
generations of halos are not much affected by the special initial
conditions of this new collisionless matter. The resulting
displacement of dark matter by stars could help resolve
observational challenges to the predicted central mass
distributions in large galaxies. And an attractive byproduct is
that the early formation of giant early-type galaxies fits a
considerable variety of observations (as reviewed in Peebles
2002).  

Second, the central parts of the most massive halos might be
expected to stop evolving as they become very much denser than
the mean density of newly collapsing halos. The point is widely
discussed, as by Navarro, Frenk \&\ White (1997; hereafter NFW) 
and in more detail by Wechsler et al. (2002). Indeed, this stable
core concept was the basis for the estimates of the redshift of
galaxy formation in Partridge \&\ Peebles (1967). Within our 
schematic model for halo formation this concept
leads us to prefer a form for the characteristic inner
halo mass density profile in the $\Lambda$CDM cosmology that is
intermediate between  
\beq 
\rho (r) = {\rho_0 \over
(r/r_s)^{3/2}[1+(r/r_s)^{3/2}]},
\label{eq:m}
\eeq
(Ghigna et al. 2000; see also Moore et al. 1999; hereafter
called the Moore form) and the NFW form
\beq
\rho (r) = 
	{\rho_0 \over (r/r_s)(1+r/r_s)^2}.
\label{eq:nfw}
\eeq 

Third, the cutoff in $\sigma_e$ or $M_e$ is a
striking phenomenon that has received theoretical attention but, 
as far as we are aware, no promising interpretation. In
particular, the widely discussed threshold for thermal
bremsstrahlung cooling (Binney 1977; Rees \&\ Ostriker 1977) does 
not apply here, because at the central halo densities presented 
by the $\Lambda$CDM model the cooling time is much shorter than
the Hubble time (see also Thoul \& Weinberg 1995). The
cosmological picture we are proposing is simple and reasonably 
well specified, and it offers an interpretation of the abundance
of massive ellipticals over some four decades in comoving number
density. This certainly is not a compelling argument for our
picture, but it does recommend close attention to the postulates.
We return to this point, and some challenges to the picture, 
in \S 3. 

\section{Computation}

\subsection{Pure CDM Halos}

We ignore dissipative settling of the baryons for the moment. We use a
modified Press-Schechter (1974) model to calculate the number density of
collapsed halos as a function of mass and redshift, and we use the
halo density profiles in equations~(\ref{eq:m}) and~(\ref{eq:nfw}) to
find the velocity dispersion at the effective radii $R_e\sim 10$~kpc
characteristic of giant elliptical galaxies.

For definiteness we adopt fixed values for most parameters. We use the
$\Lambda$CDM cosmological model, with Hubble parameter $H_0 = 70
\hbox{ km s}^{-1}\hbox{ Mpc}^{-1}$, and density parameters
$\Omega_\Lambda = 0.7$ and $\Omega _m = 0.3$, in the cosmological
constant $\Lambda$ (or dark energy that acts like $\Lambda$) and in
the sum of dark matter plus baryons, respectively. The primeval mass
density fluctuation spectrum is taken to be scale-invariant ($n=1$)
and normalized to rms mass fluctuation $\sigma _8=0.9$ in randomly
placed spheres with radius $800$ km~s$^{-1}/H_0$.

In the now standard collapse picture, a dark matter halo of total mass
$M$ at redshift $z$ that has just collapsed has virial radius (Bryan
\& Norman 1998; Barkana \& Loeb 2001)
\begin{equation}
R_v=31 \left( {M \over 10^{12} M_\odot}\right)^{1/3}
\left[{\Omega_m\over \Omega_m^z}{\Delta_c\over 18\pi^2}\right]^{-1/3}
\left({1+z\over 7}\right)^{-1}~{\rm kpc},
\end{equation}
and circular velocity
\begin{equation}
V_v=\left({GM\over R_v}\right)^{1/2}=375 \left( {M \over 10^{12}
M_\odot}\right)^{1/3} \left[{\Omega_m\over \Omega_m^z}{\Delta_c\over
18\pi^2}\right]^{1/6} \left({1+z\over 7}\right)^{1/2}~{\rm km~s^{-1}},
\end{equation}
where the density contrast at the virial radius is
$\Delta_c=18\pi^2 +82d-39d^2$, with $d\equiv \Omega_m^z-1$ and
$\Omega_m^z=\Omega_m(1+z)^3/[\Omega_m(1+z)^3+\Omega_\Lambda]$.

Our estimate of the mass function of halos is based on the
Press-Schechter (1974) model including the modification by Sheth
\& Tormen (1999; see also Sheth et al. 2001) that provides an
excellent fit to state-of-the-art N-body simulations (Jenkins et
al. 2001).

\centerline{{\vbox{\epsfxsize=8cm\epsfbox{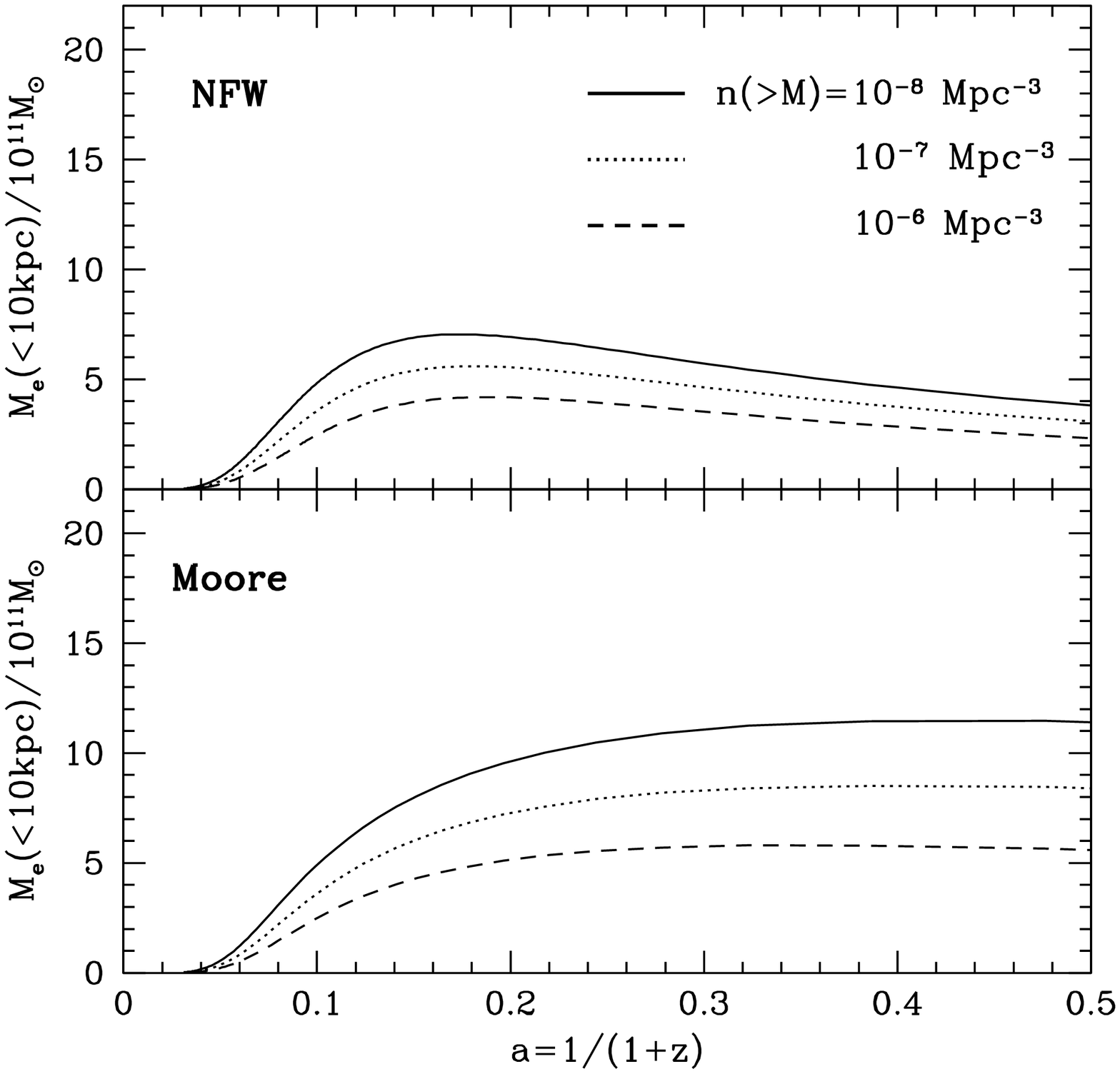}}}}
\figcaption
{The development of stable cores in CDM halos.  We show the
evolution of the mass $M_e(t)$ within a central physical radius
$R_e=10$~kpc of halos defined by a fixed value of the comoving number
density $n(>M)$. The evolution is shown as a function of the
cosmological scale factor $a=(1+z)^{-1}$. As the cores approach
stability at radii $\sim R_e$, the curves asymptote to a nearly
constant value.
\label{fig-1}
}
\vskip 0.2in

We use the Moore and NFW density profiles in equations~(\ref{eq:m})
and~(\ref{eq:nfw}) to extrapolate from $V_v$ at the virial radius to
the circular velocity at an effective radius $R_e$ for giant
ellipticals, and we divide the circular velocity by the factor
$2^{1/2}$ to estimate the line-of-sight velocity dispersion. This
is a crude approximation: our halos are not isothermal and the 
stellar velocities need not be isotropic. However, since redistribution
of the stellar mass is likely to occur (see \S 3), and some ellipticals
are known to have anisotropic velocity distributions, for the
sake of simplicity we prefer not to refine the calculation. 
In our approximations the stellar velocity dispersion at
radius $R_e$ is 
\beq 
\sigma _e =V_v\left\{ \ln [1+(cx)^{3/2}]\over
2x\ln [1+c^{3/2}]\right\}^{1/2},\qquad x = r/R_v, \eeq for the Moore
profile, and \beq \sigma_e=V_v \left\{ {\ln(1 + cx) - cx/(1 + cx)
\over 2x[\ln(1 + c) - c/(1 + c)]} \right\}^{1/2}, \eeq for the NFW
profile. In both models we choose a single value for the concentration
parameter: for NFW, $c=c_{\rm NFW}\equiv (R_v/r_s)= 4$, which is typical 
of the results from fits of the NFW profile to numerical
simulations of the more massive newly collapsed CDM halos (NFW;
Wechsler et al. 2002), and  for the corresponding Moore profile 
$c = c_{\rm NFW}/1.72$, 
which is the adjustment recommended by Klypin et al.
(2001).\footnote{Klypin et al. (2001) find that this is the
typical ratio of concentration parameters when the virial radius
and the radius at maximum circular velocity are constrained to be
the same in the two functional forms.} To avoid confusion we
remind the reader that we are considering the rarest most massive
halos that tend to have collapsed close to the redshift at which
they are identified. We suggest such halos tend to grow by the
addition of matter to the outer envelope, causing the break
radius $r_s$ to increase in rough proportion to the virial
radius, $R_v\sim cr_s$ with $c$ constant. 
 
Figure 1 shows the evolution of the mass $M_e(t)$ within a central
physical radius $R_e=10$~kpc of halos defined by a fixed value of the
comoving number density $n(>M)$.  Here and throughout the paper,
number densities are comoving and normalized to the present epoch. The
number densities belonging to the curves in the figure are in the
range of the Sloan Digital Sky Survey (SDSS) measurements for giant
early-type galaxies.  

Both models for the halo density profile, NFW in the upper panel and
Moore in the lower panel, predict similar behavior at 
$z\gtrsim 6$ because the mass distributions at radii larger than
the maximum circular velocity radius are quite similar. Both
models indicate that the mass $M_e(<10\hbox{ kpc})$ at fixed
comoving number density does not evolve much at
$z<4$. This is in line with the idea noted above that the central core
of a very massive halo stabilizes dynamically at late times, when the
core is much denser than newly collapsing halos. New mergers tend to
add mass to the outer halo envelope at impact parameters $\gg 10~{\rm
kpc}$ (due to the expansion of the universe), mostly in much smaller
halos. In the NFW form the mild decline of $M_e(<10~{\rm kpc})$
at low redshifts is a result of the smaller power law index at $x\ll
1$. This requires either that late mergers, at virial radii of a few
hundred kpc, tend to lower the densities in the inner 10~kpc of the
most massive galaxies, or that late assembly happens to produce new
galaxies with $\sigma\gtrsim 200$ km~s$^{-1}$ at $R_e\sim
10$~kpc. Since both options seem unlikely to us we conclude that
a form closer to Moore is more useful within our approximations.
We emphasize that we cannot judge which halo model would be more
useful under a better approximation to how halos form.

\vskip 0.4in
\centerline{{\vbox{\epsfxsize=8cm\epsfbox{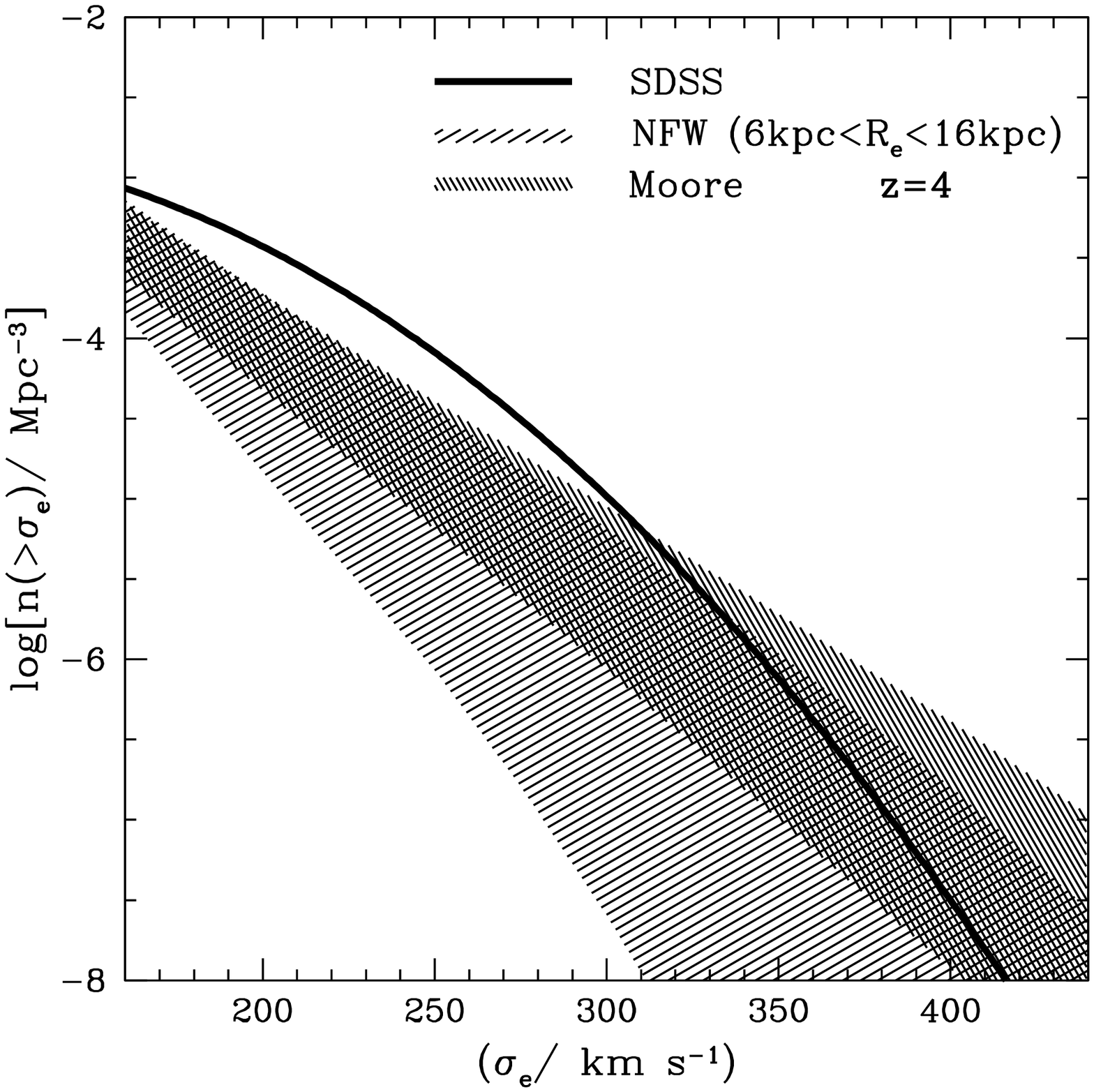}}}}
\figcaption
{Comoving number density of galaxies with a stellar velocity
dispersion above $\sigma_e$ as a function of $\sigma_e$.  The shaded
bands show the abundance predicted by populating CDM halos (having NFW
or Moore profiles) with stars at radii in the range $6~{\rm
kpc}<R_e<16~{\rm kpc}$ at $z=4$. (Similar results 
are obtained for any $z\la 4$.) The solid line shows the fitting
function derived by Sheth et al. (2002) to describe the SDSS data in
the local universe. 
\label{fig-2}
}
\vskip 0.4in

We conclude from Figure~1 that the approximate redshift at which the
assembly of the mass concentrations characteristic of giant elliptical
galaxies nears completion is 
\beq
z_f \sim 6.
\eeq  
This is close to the half-mass point in the Moore case, and just
before the peak in the NFW case.

Figure 2 shows the comoving number density of dark matter halos,
$n(>\sigma _e)$, at redshift $z=4$, as a function of the
one-dimensional velocity dispersion $\sigma _e$ computed between two
bracketing radii, $R_e=6$~kpc and $R_e=16$~kpc. These are one standard
deviation above and below the mean effective (half-light) radius
$R_e\sim 10$~kpc of giant ellipticals at $\sigma >300$ km~s$^{-1}$ in
the SDSS sample (Bernardi et al. 2001). There is a smaller spread of
values of $n(>\sigma _e)$ between the two radii in the Moore form,
because the density profile is closer to the limiting isothermal
case. A more complete computation would convolve the probability
distribution $dn/d\sigma _e$ at fixed $R_e$ with a model for the
distribution of $R_e$, and would take account of the distribution
in values of the concentration index $c$, but we leave that for
future work.

The choice of redshift in Figure 2, $z=4$, is slightly past the
characteristic epoch $z_f$ at which structure formation nears
completion in the rare massive objects we are considering, and it
is close to the redshifts reached in deep rest-frame optical
galaxy surveys (Rudnick, Rix, \&\ Franx 2001; Cimatti et al.
2002). In the  Moore model the distribution $n(>\sigma _e)$ is
not very sensitive to 
time at $z<4$. The NFW model predicts a slight decrease with
increasing time, which we are suggesting is an artifact of the
slightly too shallow inner power-law slope (within our
approximations). 

The solid line in Figure 2 shows the fitting formula for 
the measured abundance of early-type
galaxies as a function of velocity dispersion in the SDSS sample
(Sheth et al. 2002).  The standard deviation of the measurement error,
$\delta\sigma_e\sim 25$ km~s$^{-1}$, is small enough not to appreciably
broaden the steep observed drop of the distribution function at
$n(>\sigma_e)\sim 10^{-7}$~Mpc$^{-3}$.

The comparison of the SDSS data to our model depends on the inner
power-law index $\alpha\equiv -(d\log \rho/d\log x)$.  Syer \&\ White
(1998) present an elegant argument for the value of $\alpha$: if the
primeval power spectrum varies as $P(k)\propto a^2k^n$, where $a(t)$
is the cosmological expansion factor and $k$ is the comoving
wavenumber, then stable clustering indicates $\alpha = (9+3n)/(5+n)$.
Subramanian, Cen, \& Ostriker (2000) and Ricotti (2002) have checked
this relation against numerical simulations.  In the $\Lambda$CDM
model, the value of the effective index $n\equiv d\log P(k)/d\log k$
increases with mass scale.  At the wavenumbers characteristic of the
halo masses of interest, from $10^{12}M_\odot$ to $10^{15}M_\odot$,
the primeval power spectrum yields values of this index between $n\sim
-2.1$ and $-1.4$, implying $\alpha$ between $0.93$ and 1.33. Because
this range is intermediate between the NFW and Moore inner slopes, our
model should best be considered as intermediate between these two
cases in Figures 1 and 2.  It is encouraging that the data support
this intermediate regime across several orders of magnitude in galaxy
number density.  Before considering the possible significance of this
result we must deal with the loading of the dark halos by the settling
of baryons.

\subsection{Halo Loading by Star Formation}

In the SDSS sample, the ellipticals with $\sigma _e> 300$ km~s$^{-1}$
have mass-to-light ratio $M/L_{r^\star}\simeq 6$ solar units within
$R_e$ (Bernardi et al. 2001).  This is about twice that of the stars
in the Solar neighborhood.
Since the nearby stars surely are on average younger than the
populations in a giant elliptical, it seems likely that the mass
fraction in stars within the effective radius of a giant elliptical is
larger than that of the CDM component (Gerhard  et al. 2001),
\beq \eta = {M_{\rm stars}\over M_{\rm CDM}}\gtrsim 1.
\label{eq:eta}
\eeq

In the adiabatic approximation -- where the product of length and
velocity scales is conserved -- the addition of the stellar mass to
a region that contains dark mass $M$ produces final mass and
scaled radii and velocities  
\beq
M_e = (1+\eta )M, \qquad R_e = R_i/(1+\eta ),\qquad
	\sigma _e = (1+\eta )\sigma _i.
\label{eq:scaling}
\eeq
The velocity dispersion $\sigma_i$ before compression is larger
than in Figure~2 because it is computed at a larger radius,
$R_i=(1+\eta )R_e$. And the observed velocity dispersion is
larger than $\sigma _i$ by another factor $1+\eta$. 
This results in quite unacceptable velocity dispersions
unless $\eta$ is much less than unity, which does not seem
likely. 

We are not able to judge whether a more violent addition of stellar 
mass could have a less severe effect on $\sigma_e$, but the
indication from equations~(\ref{eq:eta}) and~(\ref{eq:scaling})
is that the loading of the dark halos by baryon settling could be
a serious problem for the $\Lambda$CDM model. We turn now to a 
possible remedy.

\subsection{The Attractor Hypothesis}

Numerical simulations of the growth of halos out of pure dark matter
suggest that the strongly nonlinear part of the density profile is not
very sensitive to initial conditions. A dramatic example in Navarro,
Frenk \&\ White (1996) shows a numerical simulation that evolves
through an expansion factor of just $1+z_f=5.5$. Because the initial
density fluctuations are not large this in effect significantly
truncates the small-scale initial power spectrum, yet it produces
close to standard halo density profiles. This is demonstrated in more
detail, along with the effects of other modifications of the shape of
the primeval power spectrum, by NFW and Eke et al.  (2001).

We apply this indication of a dynamical attractor effect (Syer \&\
White 1998) to the case where the small-scale mass clustering has been
increased by dissipative settling of the baryons, rather than
truncated.  Our working assumption is that stars that form prior to
the assembly of the core simply replace the dark matter that was
supposed to be there at late times. This is a conjecture: we are not
aware of any numerical checks of this case.  We note that this
conjecture could in principle explain the observed absence of a cusp
in the central dark matter distribution of nearby galaxies nad galaxy
clusters (the so-called `central cusp problem').

\section{Discussion}

Our analysis does not do justice to the precise SDSS measurements of
the abundance of early-type galaxies as a function of the stellar
velocity dispersion: within our approximations that would require
consideration of the sensitivity of the computed $n(>\sigma _e)$ to
the slope and normalization ($\sigma_8$) of the primeval power
spectrum; the distribution of values of the concentration parameter
$c$; functional forms intermediate between Moore and NFW (e.g., Power
et al.  2002); the distribution of values of galaxy effective radii
$R_e$; and the conversion from the circular velocity at $R_e$ to the
stellar velocity dispersion $\sigma_e$, which depends on a model for
how the stars populate the halo. Within the spread of possibilities
offered by all these parameters, we can only conclude from the
exploratory analysis presented here that our model seems to be capable
of accounting for the observed upper bound on the mass concentrated in
the largest galaxies.

Our model depends on the hypothesis that halo formation can erase the
effect of dissipative settling of the baryons. We are not aware of a
direct test by numerical simulations; a check would be feasible and
useful. Also open for discussion, and much more difficult to test, is
our assumption that star formation in the neighborhood of a giant
elliptical is concentrated in the dense regions that end up in or near
$R_e\sim 10$~kpc. Even if star formation were confined to dense
regions, mergers would cause diffusion of stars away from $R_e$ 
(Johnston, Sackett, \&\ Bullock 2001).  Diffusion could account for
the extended optical halos of large ellipticals (Arp \&\ Bertola
1971); numerical simulations might show whether the amount of
diffusion is acceptable at the high redshifts of formation in our
model. 

Our analysis assumes that star formation in giant ellipticals is close
to complete before their assembly at $z_f\simeq 6$. This certainly
seems consistent with the short cooling times in the central regions
of the most massive halos. And this early star formation seems to be
required in the $\Lambda$CDM model, because late star formation would
produce unacceptably large velocity dispersions in giant ellipticals,
as discussed in \S 2.2. Peacock et al. (1998) present another
consideration that leads to a similar value for $z_f$. They start from
the observed comoving number density of giant ellipticals and their
stellar velocity dispersion which they set equal to the velocity
dispersion at the virial radius, $V_v/\sqrt{2}$. Based on the
Press-Schechter mass function, they also infer a formation redshift
$z_f\sim 6$.  Although in general $\sigma_e$ is not equal to
$V_v/\sqrt{2}$, the approximation is acceptable here because $r_s \sim
10~{\rm kpc}$ at $z\sim 6$ (see Fig. 1).

There has been considerable discussion of observational constraints on
the formation redshift of giant ellipticals (Kauffmann, Charlot, \&
White 1996). In the recent deep K-band survey of Cimatti et al. (2002)
the counts of galaxies at $2\la z\la 3$ are consistent with early
formation of very luminous galaxies. This is in line with the
persuasive case by Dunlop et al. (1996) and Waddington et al. (2002)
that some giant ellipticals formed at $z_f\gtrsim 4$, and with the
evidence that many giant galaxies formed not much later than that
(Zirm, Dickinson, \& Dey 2002; Saracco et al. 2002). Other arguments
for early formation of late-type galaxies are reviewed in Peebles
(2002).

Our postulate that stars replace dark matter in a near universal form
for the net mass distribution in a giant elliptical may have some
bearing on the observation that the varying mix of baryonic and cold
dark matter as a function of radius in some ellipticals adds up to a
simple form for the total mass density, $\rho\sim r^{-2}$ (Romanowsky
\&\ Kochanek 2001; Koopmans \&\ Treu 2002; but for an exception see
Sand, Treu, \&\ Ellis, 2002; and see van Albada \&\ Sancisi 1986 for
the analog of this curious ``conspiracy'' in spiral galaxies). The
interpretation must be more complicated than a universal mass density
run, however, because the power law is steeper than Moore. The
complication may result from the rearrangement of the stellar mass
distribution by recycling of mass shed by evolving stars and from the
addition of mass by low levels of merging and accretion.  The spectra
of large early-type galaxies show evidence of ongoing star formation
(J\o rgensen 1999; Trager et al. 2000; Menanteau, Abraham, \&\ Ellis
2001), amounting to a few tens of percent of the total at $z<1$. If
this reflects the recycling of baryons in stars it might be expected
to move baryonic mass to smaller radii, making the density run
steeper, as needed. 

Merging at low redshift, driven by dynamical friction, is an important
element in semi-analytic models for galaxy formation (Cole et
al. 2000), and it is observed, as in the recent capture of a spiral by
the nearby giant elliptical Centaurus A (NGC~5128), which has
increased the mass in stars and gas in this elliptical by about 10\%
(Israel 1998). The Centaurus elliptical has many more late-type
satellites and group members (C\^ot\'e et al. 1997). But arguing
against substantial growth by accretion is the evidence that the
abundance of iron group elements relative to $\alpha$ elements is
higher in younger late-type galaxies than in ellipticals, in line with
the idea that the stars in ellipticals formed too rapidly for
appreciable enrichment of iron from type Ia supernovae (Thomas et
al. 1999; Pagel 2001). We will be following with interest constraints
from the chemistry on the amount stellar mass that merging have added
to the centers of giant ellipticals, and the effect on the central
mass density run.

Our model predicts that the most massive ellipticals reside in
very rich clusters of galaxies. It would be interesting to see
the results of a simple test, a comparison of the spatial
autocorrelation functions of massive ellipticals and of rich 
clusters with the same comoving number density. 

The proposed early formation of the giant ellipticals may help account
for the luminous quasars at $z\sim 6$ found by SDSS (Fan et
al. 2001). If these objects are radiating isotropically at the
Eddington limit then in the standard quasar model they are powered by
black holes with mass $M_{\rm BH}\ga 10^9M_\odot$, which is close to
the largest masses inferred for central compact objects in present-day
galaxies. The $\Lambda$CDM model does have initial conditions for the
formation of these massive black holes and their host galaxies at
$z\sim 6$ (Barkana \& Loeb 2002; Wyithe \& Loeb 2002). The issue for
our purpose is whether this early assembly is the dominant mode of
formation of the giant ellipticals, as we are proposing, or whether
these galaxies and their central black holes grew by a hierarchy of
mergers at redshifts well below $6$ (Haehnelt \& Kauffmann 2000). In
the latter case one might have expected that the SDSS black holes have
since grown considerably more massive than $10^9M_\odot$. That would
not naturally fit the correlation of $\sigma _e$ with $M_{\rm BH}$ in
present-day galaxies (Merritt \& Ferrarese 2001; Tremaine et al. 2002)
together with the sharp cutoff in $\sigma _e$.

Our conclusion from these considerations is that there is no serious
observational problem but instead some possible encouragement for
the idea that the giant ellipticals formed at high redshift. 

The reader may have noticed that we are arguing for early formation of
the most massive galaxies in a cosmology, $\Lambda$CDM, that usually
is associated with the formation of massive galaxies at low redshift
(see, e.g.  Figure~13 in Baugh et al. 1998). We offer two
considerations. First, Figure 1 indicates that, within commonly
accepted approximations to structure formation in this cosmology,
there is little late time addition to the mass concentrated within
10~kpc of the centers of the most massive galaxies. Perhaps the
accretion at low redshifts seen in numerical simulations of this
cosmology requires some modification of this statement, or perhaps it
requires some modification of the model that would also bring it into
agreement with the void phenomenon (Peebles 2001). Second, our early
formation scenario may apply to the giant early-type galaxies and the
late scenario to massive late-type galaxies. But the circular
velocities of spiral galaxies also show a strong upper cutoff
(Giovanelli et al.  1986, and references therein), and one would
surely hope to find a common explanation for this striking effect in
both types of galaxies.

The remarkable success of the $\Lambda$CDM cosmology in fitting
the anisotropy of the thermal cosmic background radiation (Bond
et al. 2002) argues in favor of this model as a useful
approximation to aspects of reality. We are inclined to add to the
evidence the promise of an explanation of the remarkable bound 
on mass concentrations in the most massive galaxies.

\acknowledgments 

We are indebted to Mariangela Bernardi and Ravi Sheth for kindly
sharing their SDSS data analysis with us, and to them and 
Rennan Barkana, Jerry Ostriker, Simon White, and the referee for
useful discussions. 
AL acknowledges support from the Institute for Advanced Study,
the John Simon Guggenheim Memorial Fellowship, and NSF grants 
AST-0071019, AST-0204514. PJEP acknowledges support from the NSF.

\end{document}